\documentclass{article}

\usepackage[sglblindworkshop, final]{neurips_2025}

\usepackage{graphicx}
\usepackage{bm}
\usepackage{amsmath}
\usepackage{enumitem}

\newcommand{\Mstar}{M_{\star}}
\newcommand{\Zmet}{Z}
\newcommand{\sSFR}{\mathrm{sSFR}}

\usepackage[utf8]{inputenc} 
\usepackage[T1]{fontenc}    
\usepackage{hyperref}       
\usepackage{url}            
\usepackage{booktabs}       
\usepackage{amsfonts}       
\usepackage{nicefrac}       
\usepackage{microtype}      
\usepackage{xcolor}         

\newcommand{\RR}{\mathbb{R}}

\title{Beyond Point Estimates: Benchmarking Uncertainty Quantification Methods on the AION-1 Astronomical Foundation Model}

\author{%
  Karla Tame-Narvaez\\
  Scientific Computing Division\\ 
  Fermi National Accelerator Laboratory\\ 
  Batavia, IL 60510 \\
  \texttt{karla@fnal.gov} \\
  \AND
  Aleksandra \'Ciprijanovi\'c \\
  Scientific Computing Division\\ 
  Fermi National Accelerator Laboratory\\ 
  Batavia, IL 60510 \\
  Department of Astronomy and Astrophysics\\
  University of Chicago\\ 
  Chicago, IL 60637\\
  NSF and Simons SkAI Institute\\
  Chicago, IL 60637\\
  \texttt{aleksand@fnal.gov}
  \And
  Shubhendu Trivedi \\
  Fermi National Accelerator Laboratory\\ Now at
Google DeepMind\\
  \texttt{shubhendu@csail.mit.edu}
}

\begin{document}

\maketitle

\begin{abstract}
Foundation models for astronomical surveys offer powerful learned representations that can be transferred to downstream regression tasks such as galaxy property estimation.  However, point predictions alone are insufficient for scientific inference; reliable uncertainty quantification (UQ) is essential. We compare seven UQ methods on galaxy property regression using frozen AION-1 foundation-model embeddings, predicting redshift, stellar mass, stellar-population age, gas-phase metallicity, and specific star-formation rate, from Legacy Survey photometry/imaging and DESI spectra, with PROVABGS-derived labels.  Distribution-free conformal methods achieve marginal coverage within $\sim$1\,pp of the nominal 90\% across all properties, while non-conformal baselines (Deep Ensembles, MC~Dropout) fail to calibrate reliably.  Among conformal approaches, Conformalized Quantile Regression (CQR) delivers the best coverage in the bin with the poorest model predictions. More importantly, only the Locally Valid and Discriminative (LVD) framework---particularly when operating on AION-1 embeddings---also provides finite-sample \emph{local validity}, producing intervals that adapt to each galaxy's local prediction difficulty rather than relying on marginal guarantees alone.  These results establish conformal prediction, and LVD in particular, as the preferred UQ framework for uncertainty-aware inference on foundation-model embeddings in astrophysics.
\end{abstract}

\section{Introduction}
 
The estimation of physical properties of galaxies from photometric
imaging data is a cornerstone of observational cosmology and galaxy
evolution studies.  Properties such as photometric and spectroscopic redshift, stellar
mass, stellar and gas metallicity, stellar-population age, and specific
star-formation rate encode the physical processes governing galaxy formation, from star formation and chemical enrichment to environmental quenching and the galaxy–halo connection. In recent years, deep learning approaches have demonstrated strong performance in inferring galaxy properties from heterogeneous and high-dimensional data~\cite{Gai2024,Li2025,Ayubinia2025,Rou2025}. 
Among these, \emph{foundation models}~\cite{AstroClip2024,AION}---large
neural networks pre-trained on broad, unlabeled datasets via
self-supervised objectives---have emerged as a particularly promising
paradigm.  By learning generic, transferable representations of the
input data, foundation models decouple the costly pre-training phase
from lightweight, task-specific fine-tuning, classification or regression heads, thereby
enabling rapid deployment across heterogeneous downstream tasks.
 
The AION-1 model~\cite{AION} exemplifies this strategy in
the astronomical domain. Pre-trained on multi-band stellar and galaxy data from the Legacy Survey~\cite{LegacySurveys},  Hyper Suprime-Cam
(HSC)~\cite{HSC}, Sloan Digital Sky Survey (SDSS)~\cite{SDSS}, the Dark Energy Spectroscopic Instrument (DESI)~\cite{DESI}, and Gaia~\cite{Gaia}, AION-1 produces high-dimensional embeddings which integrates complex information from heterogeneous imaging, spectroscopic, and scalar data. Training is done using a two-stage architecture: modality-specific tokenization followed by transformer-based
masked modeling of cross-modal token sequences. AION-1 achieves strong results on a broad suite of downstream tasks, such as galaxy and stellar property estimation, galaxy morphology classification, similarity search, galaxy image segmentation, and spectral super-resolution. 

In this work we will focus on downstream regression of galaxy properties from AION-1 embeddings. 

However, point estimates presented in AION-1 paper are insufficient for scientific inference.  Robust \emph{uncertainty quantification} (UQ) is crucial for propagating measurement and modeling errors, identifying out-of-distribution inputs and distribution shift, and ensuring that both individual predictions and population-level inferences remain reliable and scientifically robust. Despite its importance, UQ in deep learning remains only partially understood, and for large, black-box foundation models—with their complex, high-dimensional embeddings—it is especially challenging and largely unexplored in astrophysics and science in general. In this paper, we conduct a direct comparison of seven UQ methods---LDV and LVD-MAD~\cite{LVD}, VanillaSplit~\cite{Vovk2005}, MADSplit~\cite{Lei2016,MADSplit}, CQR~\cite{Romano2019CQR}, Deep Ensembles~\cite{Lakshminarayanan2017}, MC Dropout~\cite{Gal2016Dropout}---applied to the same AION-1 backbone and the same downstream regression task as in Table 1 from~\cite{AION}. We measure the marginal, tail, and worst-predicted coverage, as well as efficiency (mean prediction interval width) for each UQ method, and discuss problems and potential pitfalls related to some of these commonly used UQ methods. 
 
\section{Data}

We use galaxy samples from three AION-1 surveys: photometry ($\{g,r,i,z,y\}$ fluxes) and $\{g,r,z\}$ imaging from the DESI Legacy Imaging Surveys, $3600\text{--}9800\,\textup{\AA}$ DESI spectra, and PROVABGS SED-fitting posteriors providing labels for redshift, stellar mass, age, metallicity, and specific star-formation rate\footnote{We note that these labels are themselves derived quantities 
from SED-fitting rather than direct observables, and therefore 
carry their own modeling uncertainties; our UQ evaluation measures 
calibration with respect to these derived labels, which in some cases might differ from true physical values.}. We leverage tokens from the pre-trained AION-1 model, but train a task-specific regressor using these tokens and PROVABGS labels. The dataset consists of 50{,}000 galaxies (out of 120{,}000 total) for training and 11{,}362 for testing, ensuring no overlap with AION-1 pre-training via object IDs. For downstream UQ tasks, the training set is further split 70\%/30\% into train/calibration. Final dataset sizes are 34{,}844 objects for training, 14{,}934 for calibration and 11{,}362 for testing.

Each dataset is tokenized following the modality-specific procedures described in the AION-1 framework. The resulting token sequences are concatenated and passed through the frozen AION-1 encoder, which produces a contextualized 768-dimensional embedding vector for each galaxy via attentive pooling.  This configuration corresponds to the photometry\,+\,imaging\,+\,spectra (Ph+Im+Sp) input mode of AION-1, which was shown to yield the highest $R^2$ across all five targets \footnote{Throughout this test study we use the AION-B variant (300\,M parameters), the smallest member of the AION-1 model family---which also includes AION-L (800\,M) and AION-XL (3\,B).}. The encoder weights remain frozen throughout the study; only the downstream regression head and UQ-specific components are optimized. This design ensures that all UQ methods operate on an identical input representation and that the comparison isolates the effect of the UQ strategy itself. 

\section{Regression Network Architecture}

For each galaxy we regress five physical properties: redshift~$z$, stellar mass~$\log\Mstar$, mass-weighted stellar-population age~$t_{\mathrm{age}}$, mass-weighted gas-phase metallicity~$\log Z$, and specific star-formation
rate~$\log\sSFR$. The regression architecture used in this study is a three-layer multilayer perceptron
(MLP), which maps the 768-dimensional AION-1 embeddings to a scalar prediction for each target property. The network consists of two hidden layers of 256 units each with Gaussian Error Linear Unit (GELU) activations, followed by a single linear output neuron. The second hidden-layer activations $\bm{h}_2 \in \RR^{256}$ are returned as
intermediate embeddings for use in downstream UQ methods.
A separate MLP is trained independently for each of the five target properties.  Each model is optimized using Adam with a learning rate of $10^{-3}$ and weight decay of $10^{-5}$, paired
with a cosine annealing learning rate schedule over
$T_{\max} = 100$ cycles, for a total of 500 epochs minimizing the MSE loss. After training, the model is evaluated on the held-out test set and the coefficient of determination~$R^2$ is recorded for each target property as a point-estimate baseline. Our model produces similar $R^2$ values as in AION-1 paper, namely $\{z, \log\Mstar, t_\mathrm{age}$, $\log\Zmet, \log\sSFR\}=\{0.98,  0.89,  0.85, 0.63, 0.86\}$. The code used in this work will be shared on~\href{ https://github.com/..............}{Github}, after paper is accepted.

\section{UQ Methods}

We compare seven UQ methods and it's variants.  The \emph{Locally
Valid and Discriminative} (LVD) framework~\cite{LVD} learns a kernel
in a reduced feature space to estimate conditional nonconformity
quantiles, yielding prediction intervals with approximate
\emph{per-input} validity.  We test two embedding spaces:
\textbf{LCNet}, where the kernel operates on the 256-dimensional MLP
hidden-layer activations~$\bm{h}_2$, and \textbf{LCAion}, where it operates on the raw 768-dimensional AION-1 embeddings.   
Formally, local validity requires that the prediction interval covers the true value with probability at least $1-\alpha$ within any local neighborhood of the input, not just on average across all inputs; this is strictly stronger than marginal validity, which only guarantees the nominal coverage rate when averaged over the entire test distribution.
\textbf{LVD-MAD} extends LVD by normalizing nonconformity scores with a learned
residual-magnitude predictor, producing scale-adaptive intervals; we again have two variants: \textbf{LCMADNet} (kernel and residual predictor on~$\bm{h}_2$) and \textbf{LCMADAion} (kernel and residual on AION-1 embeddings).  For the MAD variants, the residual predictor is trained on the training ser and calibration is performed on the
held-out calibration set, preventing data leakage. For all LVD variants, the kernel is a metric-learned projection to 5 dimensions (500 iterations, learning rate $10^{-3}$, early stopping after 50 non-improving steps, with L2 normalisation); for the MAD variants, the residual magnitude predictor is a two-layer MLP (hidden size 32 for \textbf{LCMADNet}, 64 for \textbf{LCMADAion}) with ReLU activations and a Softplus output, trained for 500 steps with Adam at $10^{-3}$. The runtime for all four LVD variants is approximately 30min in a 40 GB GPU A-100.

\textbf{VanillaSplit}~\cite{Vovk2005} is standard split conformal
prediction with a single global nonconformity quantile, producing
constant-width intervals with marginal coverage guarantees.
\textbf{MADSplit}~\cite{Lei2016,MADSplit} is its MAD-normalized variant, scaling scores by a
learned residual predictor trained on the training set to
adapt interval width to local prediction difficulty.  \textbf{CQR}
(Conformalized Quantile Regression)~\cite{Romano2019CQR} trains
separate networks for the $\alpha/2$ and $1{-}\alpha/2$ conditional
quantiles on training set, then applies a conformal
correction on the calibration set to restore finite-sample marginal
coverage.

The two non-conformal baselines are \textbf{DE} (Deep
Ensembles)~\cite{Lakshminarayanan2017}, which trains $M{=}5$
independently initialized MLPs (with MSE loss) and constructs Gaussian
intervals from the inter-model predictions of the mean and standard deviation, and
\textbf{MC~Dropout}~\cite{Gal2016Dropout}, which injects dropout
($p{=}0.1$) into the pre-trained regressor at inference time and
estimates predictive mean and variance from $T{=}100$ stochastic
forward passes.  Both capture only \emph{epistemic} uncertainty---DE
through inter-model disagreement, MC~Dropout through dropout-induced
variance---with no learned aleatoric term.

Conformal methods provide distribution-free, finite-sample coverage guarantees, making them well suited to astrophysical settings with heterogeneous data, complex noise, and simulation--data mismatch. They are increasingly used for reliable uncertainty intervals in tasks such as redshift estimation and stellar population inference. In contrast, non-conformal approaches (e.g., Deep Ensembles and MC~Dropout) offer richer predictive uncertainty representations and integrate naturally with inference pipelines, but rely on model assumptions and lack formal coverage guarantees, often leading to miscalibration under distribution shift. While conformal methods ensure marginal validity, they may suffer from conditional miscalibration; methods such as MADSplit, CQR, and LVD address this through input-dependent or locally adaptive calibration, improving reliability in high-dimensional regimes.

\begin{figure}[h!]
\begin{center}
\includegraphics[width=1\linewidth]{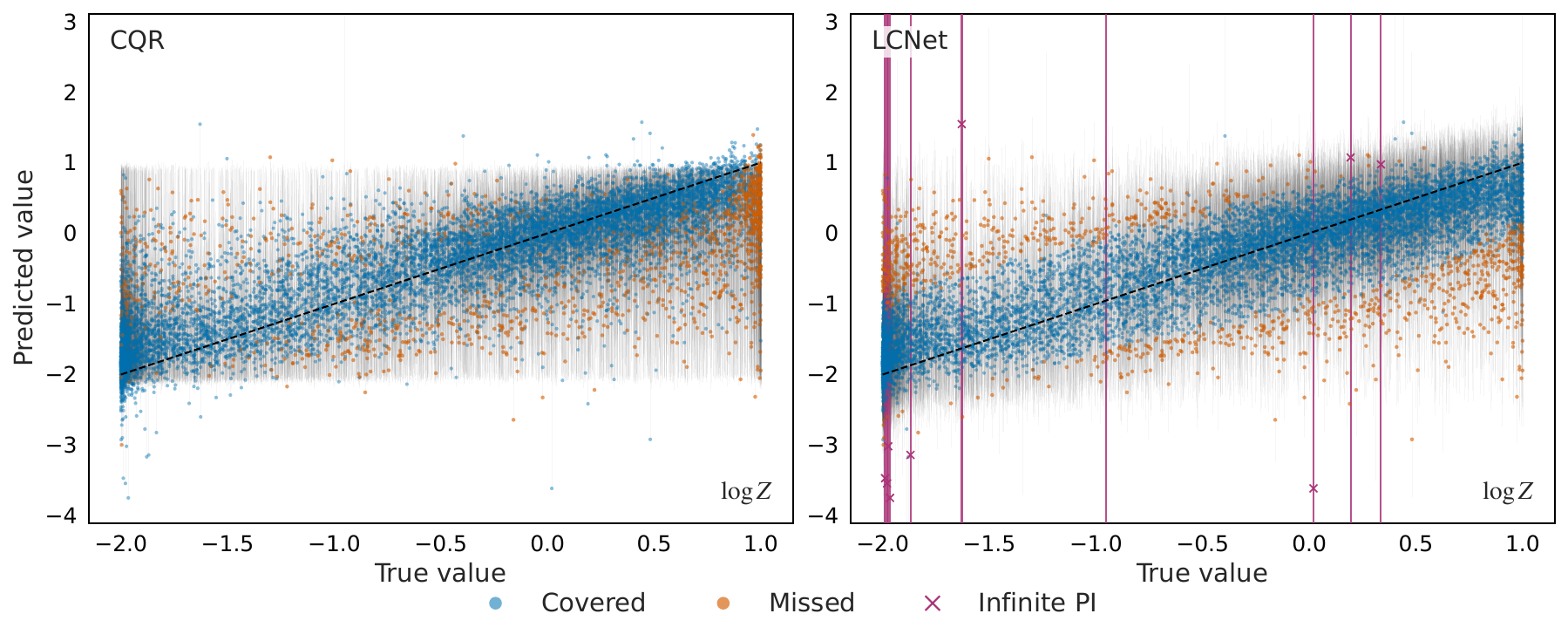}
\caption{Predicted vs. True $\log Z$ with $90\%$ prediction intervals for CQR (left) and LCNet (right). Both achieve similar marginal coverage ($\sim90\%$), but CQR produces more uniform intervals, while LCNet adapts interval widths to local prediction difficulty. A small number (10 cases) of LCNet intervals are unbounded (which is actually a desirable behaviour, as it can be used to discover outliers and anomalies).}
\label{fig:cqr_vs_lvd}
\label{fig:comp}
\end{center}
\end{figure}

\section{Results and Conclusion}

We evaluate coverage at three complementary levels: over the full test set (marginal coverage), in the $10\%$ upper and lower tails (tail coverage), and on the $10\%$ worst-predicted points (worst-predicted bin coverage). As expected, all conformal methods---including the LVD variants, VanillaSplit, MADSplit, and CQR---achieve similar $\sim$90\% marginal coverage (Table~\ref{tab:coverage_comparison}), consistent with their distribution-free guarantees.

In the tails, LVD variants remain competitive, with LCMADAion and LCAion generally achieving the most consistent performance, indicating that operating in the richer AION embedding space enhances robustness near the edges or in low-density regions. This advantage likely reflects structure in the $\mathbb{R}^{768}$ AION embedding that the downstream MLP discards during compression to $\mathbb{R}^{256}$, but which LVD's kernel can still exploit for local calibration. In contrast, the non-conformal baselines fail to calibrate properly: DE systematically undercovers (12--38\%) with overly narrow intervals, while MC~Dropout exhibits strong overcoverage (57--99\%) with intervals that are $2$--$3\times$ wider.

Clear differences emerge in more challenging regimes. In the worst-predicted bin, CQR attains the highest coverage (44--66\%), followed by the LVD variants, where LCMADAion (22--42\%) and LCMADNet (9--40\%) consistently outperform their non-MAD counterparts, LCAion (12--33\%) and LCNet (10--20\%). This demonstrates the importance of scale-adaptive normalization in capturing heteroscedasticity and improving local calibration. Across all targets, LVD methods maintain non-trivial coverage, while VanillaSplit collapses to $\sim$0\% in the hardest regions and MADSplit, although better (7-41\%), shows less consistency.

Among well-calibrated methods, the LVD variants achieve the best efficiency, producing the tightest intervals while preserving coverage. Although CQR shows stronger performance in marginal and worst-bin metrics, it guarantees only marginal validity. In contrast, LVD provides approximate conditional coverage through local, kernel-based calibration, ensuring reliability at the level of individual predictions (see Fig.~\ref{fig:comp}). This makes LVD---particularly the MAD and AION variants---a more principled choice when per-input validity and robustness in complex, high-dimensional regimes are required. All experiments use the AION-B (300M parameters) variant; whether the LVD-on-embedding advantage persists at the AION-L (800M parameters) and AION-XL (3B parameters) scales is left to a follow-up study. We note that the coverage figures in Table~\ref{tab:coverage_comparison} do not include uncertainty estimates; bootstrap or repeated-calibration-split error bars will be reported in a more thorough follow-up study, to better understand the stability of each UQ method.

\begin{table*}[ht!]
    \centering
    \caption{Marginal coverage, tail coverage, worst-predicted bin coverage, and efficiency (mean prediction interval width) for each UQ method at nominal level $1-\alpha=0.90$.}
    \label{tab:coverage_comparison}
    \setlength{\tabcolsep}{3.5pt}
    \scriptsize
    \renewcommand{\arraystretch}{1.15}

    \setlength{\tabcolsep}{3.5pt}
    \begin{tabular}{l ccccccccc}
        & \rotatebox{70}{LCNet}
        & \rotatebox{70}{LCAion}
        & \rotatebox{70}{LCMADNet}
        & \rotatebox{70}{LCMADAion}
        & \rotatebox{70}{VanillaSplit}
        & \rotatebox{70}{MADSplit}
        & \rotatebox{70}{CQR}
        & \rotatebox{70}{DE}
        & \rotatebox{70}{MC Dropout} \\
        \midrule
        \multicolumn{9}{c}{\textbf{Marginal Coverage (\%)}} \\
        \midrule
        $z$                        & 89.64 & 91.42 & 90.07 & 91.43 & 89.48 & 89.49 & 90.04 & 38.14 & 99.18 \\
        $\log\Mstar$               & 88.76 & 90.09 & 88.82 & 89.71 & 89.54 & 89.86 & 89.78 & 11.68 & 99.78 \\
        $t_\mathrm{age}$           & 90.68 & 90.55 & 92.55 & 91.11 & 90.07 & 90.83 & 90.77 & 21.62 & 97.48  \\
        $\log\Zmet$                & 90.51 & 89.06 & 88.37 & 87.91 & 89.10 & 88.43 & 88.92 & 18.93 & 56.72 \\
        $\log\sSFR$                & 89.42 & 91.16 & 89.42 & 89.59 & 89.61 & 89.59 & 90.26 & 22.53 & 99.38  \\
        \midrule
        \multicolumn{9}{c}{\textbf{Tail Coverage (\%)}} \\
        \midrule
        $z$                        & 84.33 & 87.43 & 83.85 & 87.70 & 75.62 & 77.99 & 89.83 & 34.60 & 98.37 \\
        $\log\Mstar$               & 79.89 & 79.34 & 79.75 & 83.41 & 75.31 & 83.14 & 83.10 & 9.860 & 99.16 \\
        $t_\mathrm{age}$           & 88.32 & 87.29 & 91.18 & 87.60 & 86.40 & 89.44 & 87.06 & 29.09 & 94.94 \\
        $\log\Zmet$                & 86.34 & 84.05 & 88.11 & 85.51 & 83.23 & 86.49 & 75.88 & 13.82 & 53.96  \\
        $\log\sSFR$                & 79.62 & 83.65 & 79.49 & 84.86 & 78.52 & 84.15 & 83.93 & 22.27 & 97.71  \\
        \midrule
        \multicolumn{9}{c}{\textbf{Worst-Predicted Bin Coverage (\%)}} \\
        \midrule

        $z$                        & 18.75 & 31.77 & 20.95 & 34.16 & 0.000 & 7.390 & 65.76 & 0.000 & 91.81 \\
        $\log\Mstar$               & 10.04 & 14.69 & 9.510 & 39.03 & 0.000 & 38.47 & 48.86 & 0.000 & 97.80 \\
        $t_\mathrm{age}$           & 20.19 & 16.95 & 40.48 & 22.42 & 0.700 & 27.82 & 43.93 & 0.000 & 74.82 \\
        $\log\Zmet$                & 17.27 & 12.46 & 37.71 & 28.36 & 0.000 & 26.58 & 52.73 & 0.000 & 6.16 \\
        $\log\sSFR$                & 11.88 & 32.97 & 12.59 & 41.78 & 0.000 & 41.20 & 48.42 & 0.000 & 93.84 \\
        \midrule
        \multicolumn{9}{c}{\textbf{Efficiency}} \\
        \midrule
    
        $z$                        & 0.039 & 0.041 & 0.041 & 0.041 & 0.040 & 0.040 & 0.054 & 0.016 & 0.109  \\
        $\log\Mstar$               & 0.563 & 0.575 & 0.563 & 0.554 & 0.594 & 0.558 & 0.667 & 0.060 & 2.173  \\
        $t_\mathrm{age}$           & 0.142 & 0.141 & 0.154 & 0.142 & 0.143 & 0.141 & 0.153 & 0.022 & 0.228  \\
        $\log\Zmet$                & 1.695 & 1.681 & 1.523 & 1.574 & 1.791 & 1.617 & 1.571 & 0.217 & 0.757  \\
        $\log\sSFR$                & 0.705 & 0.708 & 0.702 & 0.673 & 0.725 & 0.677 & 0.711 & 0.125 & 2.055  \\
        \bottomrule
    \end{tabular}
\end{table*}

\begin{ack}
This work was produced by Fermi Forward Discovery Group, LLC under Contract No. 89243024CSC000002 with the U.S. Department of Energy, Office of Science, Office of High Energy Physics. The United States Government retains, and the publisher, by accepting the work for publication, acknowledges that the United States Government retains a non-exclusive, paid-up, irrevocable, world-wide license to publish or reproduce the published form of this work, or allow others to do so, for United States Government purposes. The Department of Energy will provide public access to these results of federally sponsored research in accordance with the DOE Public Access Plan
(\url{http://energy.gov/downloads/doe-public-access-plan}).
The work of K.T. and A.C. is supported by DOE Grant DE-SCL0000092 (HEP AmSC IDA Pilot: AI Universe).
\end{ack}

\clearpage
\bibliography{bibliography}
\bibliographystyle{plain}


\end{document}